\documentclass[10pt,letterpaper]{article}
\usepackage{opex3}

\begin{document}

\title{Optimal light harvesting structures at optical and infrared frequencies}

\author{F. Villate-Gu\'io$^1$, F. L\'opez-Tejeira$^2$, F.J. Garc\'ia-Vidal$^3$, L. Mart\'in-Moreno$^1$, and F. de Le\'on-P\'erez$^{1,4}$}

\address{$^1$Instituto de Ciencia de Materiales de Arag\'on and Departamento de F\'isica de la Materia Condensada, CSIC-Universidad de Zaragoza, E-50009 Zaragoza, Spain}
\email{lmm@unizar.es}
\address{$^2$Instituto de Estructura de la Materia (IEM-CSIC), Consejo Superior de Investigaciones Cient\'ificas, Serrano 121, 28006 Madrid, Spain}

\address{$^3$Departamento de F\'isica Te\'orica de la Materia Condensada, Universidad Aut\'onoma de Madrid, E-28049 Madrid, Spain}

\address{$^4$Centro Universitario de la Defensa de Zaragoza, Ctra. de Huesca s/n, E-50090 Zaragoza, Spain}

\email{fdlp@unizar.es} 



\begin{abstract}
One-dimensional light harvesting structures with a realistic geometry nano-patterned on an opaque metallic film are optimized to render high transmission efficiencies at optical and infrared frequencies. Simple design rules are developed for the particular case of a slit-groove array with a given number of grooves that are symmetrically distributed with respect to a central slit. These rules take advantage of the hybridization of Fabry-Perot modes in the slit and surface modes of the corrugated metal surface. Same design rules apply for optical and infrared frequencies. The parameter space of the groove array is also examined with a conjugate gradient optimization algorithm that used as a seed the geometries optimized following physical intuition.  Both uniform and nonuniform groove arrays are considered. The largest transmission enhancement, with respect to a uniform array, is obtained for a chirped groove profile. Such enhancement is a function of the wavelength. It decreases from 39 \% in the optical part of the spectrum to 15 \% at the long wavelength infrared. 
\end{abstract}




\section{Introduction}
\begin{figure}[htbp]
 \centering
\includegraphics[width=12cm]{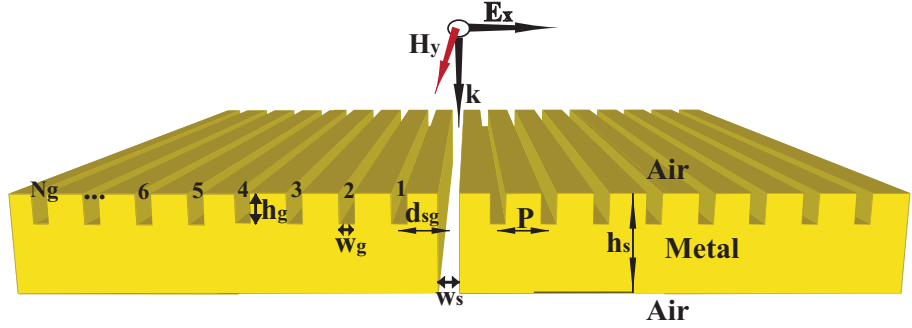}
\caption{(Color). Schematic representation of the slit-groove array on a free-standing gold film of thickness $h_s$. The  slit of width $w_s$ is surrounded by $2N_g$ grooves (width $w_g$, depth $h_g$, and period $P$). The distance from  the slit to the first groove, $d_{sg}$, is in principle different from $P$.}
\label{fig:scheme}
\end{figure}

Exciting new developments in optoelectronics and chemical sensing are based on a single subwavelength aperture milled in an opaque metal film, which surface is sculpted at the scale of the wavelength \cite{GenetN07,FJRMP10}.  In systems like the slit-groove array (SGA) sketched in Fig. \ref{fig:scheme}, the surface corrugation acts like an antenna to couple the incident light into surface modes that squeeze the EM energy into the aperture, leading to high transmission efficiencies at a narrow spectral range. Surface plasmon polaritons (SPPs), modified by the metal corrugation, are responsible for light harvesting in the optical regime; while in the THz regime, where metals behave as perfect electric conductors (PEC) and SPP modes no longer exist, surface modes are induced by the geometry \cite{PendryS04}.

 Tuning the optical response of single apertures surrounded by periodic grooves are of particular interest in the context of spectral and polarizing imaging  \cite{ThioOL01,FJPRL03,IshiJJAP05,DunbarAPL09}. Transmission resonances at a given wavelength can be controlled by adjusting the size and geometry of the SGA. Its  optical response  has been optimized by varying groove depth \cite{ThioOL01}, periodicity \cite{FJPRL03}, 
aperture size \cite{ThioNT02}, metal thickness \cite{DegironOE04}, and number of grooves \cite{JanssenPRL07}. Such studies are not limited to uniform and periodic groove arrays, where all grooves are identical and equidistant, but also consider nonuniform and not periodic structures \cite{ShiAPL07,LuJOSAB07,LiAPL08,LauxNP08,SolOE11}. 

The correlation between geometrical parameters for an efficient light harvesting process (LHP) has been discussed in several of the previous references. Authors of Ref. \cite{FJPRL03} have pointed out the role played by groove cavity modes, waveguide (Fabry-Perot) modes in the slit, and the in phase groove re-emission (controlled by the period of the groove array).  They employed the PEC approximation, which has only indicative values at optical frequencies. Janssen {\it et al.} refined the numerical exploration of the array geometry explaining how to squeeze additional light into a slit in a real metal by increasing the number of grooves but reducing the size of each groove at the same time \cite{JanssenPRL07}.

The aforementioned studies are limited to some relevant geometrical parameters and often to a fixed wavelength. Moreover, whereas such studies are mainly focused on optical, THz and microwave frequencies, less attention has been paid to the infrared (IR) regime, for which one can envisage potential applications in spectroscopic chemical sensing  \cite{Suetaka,CoeARPC08}, bio-sensing \cite{LeeOE11,TetzOL06,YanikPNAS11}, and light harvesting technologies, see Ref. \cite{RossNP12} for a recent review.  In the present paper, we go one step further and study a broader area of the parameter space and a larger spectral range that extends from optical to IR frequencies.

Our first goal is to identify simple design rules for optimal light transmittance by the SGA shown in Fig. \ref{fig:scheme}. Such rules are developed from the physical mechanisms responsible for squeezing light into the central aperture. Our second goal is to further optimize the light harvesting process  with help of a conjugate gradient (CG) maximization algorithm \cite{NRF77} that scrutinizes the parameter space of the groove array. Seeds for the CG algorithm are taken from simulations based on physical intuition. We shall consider uniform and periodic groove arrays as well as nonuniform and not periodic ones, which further enhances the light transmission through the slit.  Notice that Ref. \cite{LiAPL08} has already explored numerically  nonuniform SGAs but only in the microwave regime. Moreover, a remarkable difference with previous works is that we optimized the LHP for a fixed size of the system. Thus, we perform a full optimization for a given number of grooves and a given wavelength. Fixed size optimizations are 	more fruitful for practical applications, like optical detectors, which usually have a given pixel size and its fabrication is limited by several geometrical constrains.
  
Calculations are done in the framework of the coupled-mode method (CMM) \cite{FJRMP10} with  surface impedance boundary conditions \cite{jackson}. Such approximate boundary conditions are applied not only at the horizontal air/metal interfaces but also at the vertical walls of slit and grooves. This model nicely reproduces experimental results \cite{DunbarAPL09,FLTNP07,FLTNJP08}.  

Despite the intensive optimization presented here, a more efficient LHP  can be achieved with slight variations of the geometry of Fig. \ref{fig:scheme}. Such kind of improvements have been largely considered in the literature, e.g. the inclusion of Bragg reflectors \cite{IshiharaJJAP05}, the change of the shape of either the aperture \cite{IshiharaAPL06} or the grooves \cite{IshiJJAP05}, or the combination of single apertures perforated on a metal film with dielectric surface
gratings \cite{LinOE06}; see Ref. \cite{FJRMP10} for a comprehensive review. The study of such modified geometries exceeds the scope of the present paper. However, we hope that the present discussion could motivate further theoretical and experimental studies on that direction.

The paper is organized as follows. The coupled-mode method is briefly described in the next section. Section \ref{sec:regular} studies	 a uniform SGA. Its optical response is optimized first by a procedure based on physical intuition and next with the CG algorithm. Both wavelength and size of the system are varied in the optimization process. Section \ref{sec:nonregular} considers a nonuniform SGA of a given size as a function of the wavelength. At the end of the paper our main conclusions are summarized.

\section{Theoretical framework: coupled-mode method}
We study the system represented in Fig, \ref{fig:scheme}. It consists in a slit of width $w_s$ perforated in a free-standing gold layer of thickness $h_s$. The front metal surface is corrugated with a set of $2N_g$ grooves (of width $w_g$ and depth $h_g$) symmetrically distributed with respect to the central slit. The groove period is $P$. The distance from  the slit to the first groove, $d_{sg}$, could be in principle different from $P$. We assume hereafter that the SGA is illuminated by a plane wave which electric field is in the plane of the surface and perpendicular to the slit direction. The dielectric constant of gold is taken from Ref. \cite{JohnsonPRB72} and \cite{Palik} for optical and infrared frequencies, respectively.

We compute the intensity of the light radiated to the farfield and normalize it to the intensity of the light incident on the area of the slit.  The resulting normalized-to-area transmittance, called $\eta$ from now on, accounts for the efficiency of the light harvesting process: $\eta$ is of the order of 1 for a single slit, whereas it could become one or two orders of magnitude larger when the groove array squeezes additional light to the central slit. 

Maxwell's equations are solved self-consistently using the coupled-mode method \cite{FJRMP10}. It is based on a convenient representation of the EM fields. Above and below the metal film shown in Fig. \ref{fig:scheme}, the fields are expanded into an infinite set of plane waves with both p- and s-polarizations. Inside slit and grooves the most natural basis is a set of planar waveguide modes \cite{stratton}.  Convergence is fast achieved with a small number of such modes \cite{FLTAPA07}. The parallel components of the fields are matched at the metal/dielectric interface using surface impedance boundary conditions (SIBCs) \cite{jackson}. Although SIBCs neglect the tunneling of EM energy between the two metal surfaces, this effect is not relevant for a metal thickness larger than a few skin depths. SIBCs are also used at the lateral walls of slit and grooves. After matching the fields at the interface  we arrive to a linear system of tight binding-like equations, than can be easily solved.  We direct the interested reader to Ref.  \cite{FdLPNJP08} for the expressions of the full multimode formalism as well as their derivation, and to Ref. \cite{FLTAPA07} for its application to one-dimensional systems.


\section{Uniform slit-groove array}
\label{sec:regular}

\begin{figure}[htbp]
 \centering
\includegraphics[width=7cm]{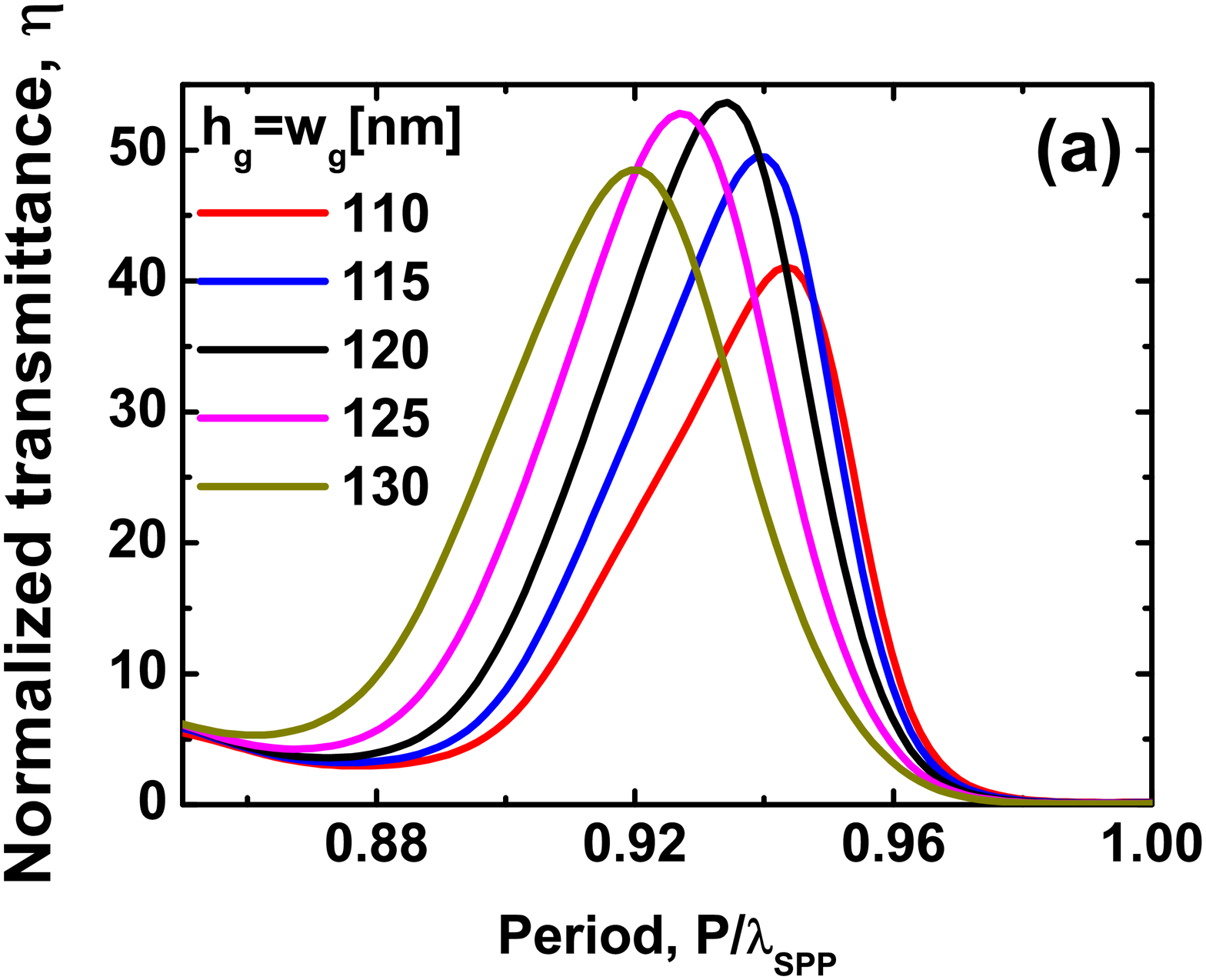}
\includegraphics[width=7cm]{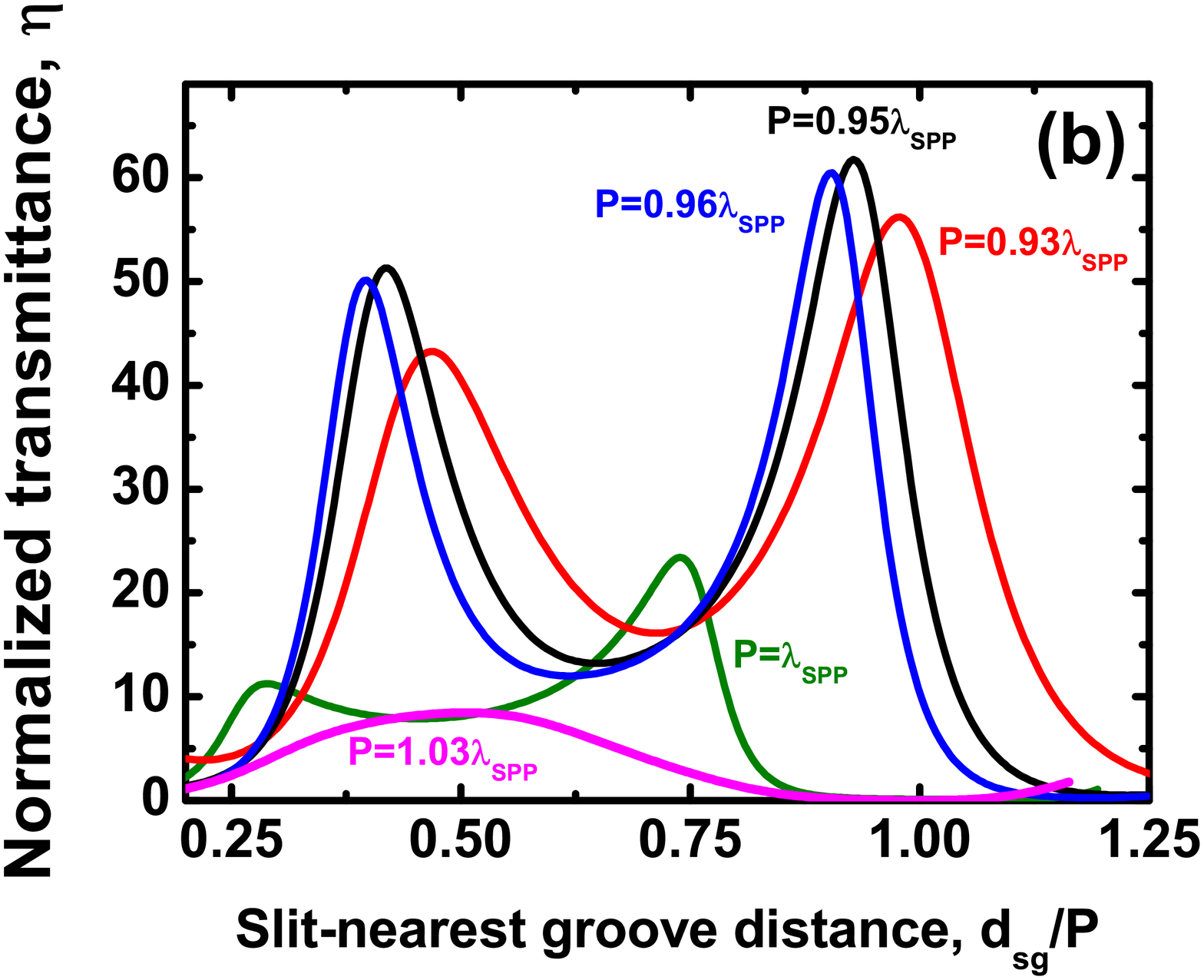}
\includegraphics[width=7cm]{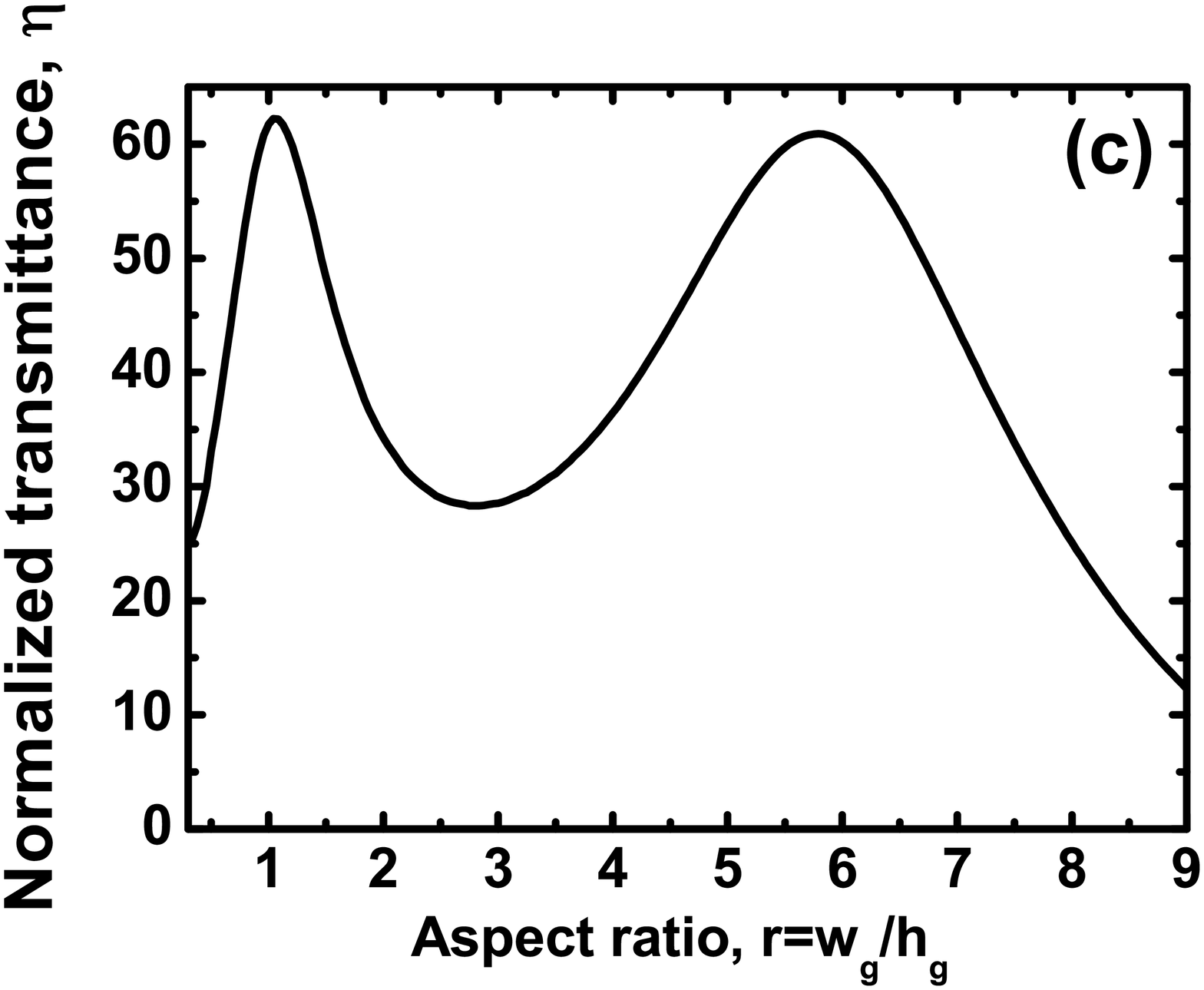}
\caption{(Color). (a) Normalized-to-area transmittance ($\eta$) for a SGA as a function of the groove period P for groove depth increasing from  $h_g$=110 nm to 130 nm at $\lambda=1.35$ $\mu$m. P is normalized to $\lambda_{spp}=1.33$ $\mu$m. (b)  $\eta$ as a function of the slit-first groove distance $d_{sg}$ for several values of $P/\lambda_{spp}$ taken from the black curve of (a) ($h_g=w_g=125$ nm). (c) $\eta$ versus the aspect radio ($r=w_g/h_g$) for the optimal geometry of (b): $P=0.95 \lambda_{spp}$ and $d_{sg}=0.93 P$.}
 \label{fig:regular}
\end{figure}

We consider first a uniform SGA, where all grooves are identical. Section \ref{subsec:phys} describes the physical mechanisms underneath in the light harvesting process and how to combine such mechanisms for having an enhanced transmittance.  The optimal geometry obtained in this way is used in sections \ref{subsec:CGvslambda} and \ref{subsec:CGvsNg} as a seed for the conjugate gradient maximization method. In section \ref{subsec:CGvslambda} the SGA is optimized as a function of the wavelength for a fixed number of grooves, while  in section \ref{subsec:CGvsNg} $\lambda$ is keep constant and $N_g$ is allowed to vary.

\subsection{Optimization following physical intuition}
\label{subsec:phys}

The normalized-to-area transmittance 	of a SGA can be enhanced when a Fabry-Perot mode is excited inside the slit or when the geometry of the groove array supports groove cavity modes. For a given wavelength, the Fabry-Perot mode can be tuned by metal thickness $h_s$ (see \cite{FJRMP10} and references therein), while the groove cavity mode of a groove array is a function of the groove depth and period \cite{FLTPRB05}. However, the largest transmittance is obtained when the Fabry-Perot mode of the slit is located at the same spectral position than the groove cavity mode, as pointed out in Ref. \cite{FJPRL03}. We find that this recipe applies not only for optical frequencies but also for the IR part of the spectrum.  Fig. \ref{fig:regular}(a) illustrates this behavior for $\lambda=1.35$ $\mu$m and $N_g=10$ grooves. The Fabry-Perot peak for a single slit is centered at  $\lambda=1.35$ $\mu$m for a film thickness of $h_s=360$ nm.  Despite  the intensity of the Fabry-Perot mode changes monotonously with the aperture size $w_s$ of a sub-wavelength slit \cite{FJRMP10,HarringtonAP80}, choosing the value of $w_s$ implies a compromise between a high normalized transmittance $\eta$ that decreases with $w_s$ and a high total transmittance T that increases with $w_s$.  Similar trends are found for a SGA. We therefore keep constant the value of $w_s$. We use $w_s=100$ nm that gives $\eta=1.8$ for a single slit at $\lambda=1.35$ $\mu$m. 

The groove cavity mode coincides with the Fabry-Perot mode for $P=0.93\lambda_{spp}$ and $h_g=120$ nm, leading to the largest $\eta=54$ in the black line of Fig. \ref{fig:regular}(a). P is normalized by the SPP wavelength ($\lambda_{spp}=1.33$ $\mu$m) for the sake of convenience. $P<\lambda_{spp}$ due to penetration of the field in the grooves that enlarges the optical path of the SPPs. SPPs are emitted in phase when $\lambda_{spp}$ is equal to the sum of $P$ and the extra optical path provided by the grooves. We assume for the moment that $w_g=h_g$ and $d_{sg}=P$. These constrains are relaxed in what follows.

Fabry-Perot and groove cavity modes are in principle hybridized in a SGA, but roughly speaking we could say that the interaction between slit and groove array can be controlled by the distance from the slit to its nearest groove, $d_{sg}$. Fig. \ref{fig:regular}(b) renders $\eta$ as a function of $d_{sg}$ (normalized by $P$) for several values of $P/\lambda_{spp}$. The groove pitch increases from the value $P=0.95 \lambda_{spp}$, at which $\eta$ has a maximum in   Fig. \ref{fig:regular} (a), to the value $P=1.03 \lambda_{spp}$, at which $\eta$ has a  minimum. Fig. \ref{fig:regular} (b) shows a double-peak when $P<\lambda_{spp}$: one peak at  $d_{sg} \approx 0.4 P$ and another at $d_{sg} \approx 0.95 P$. The double-peak is shifted to smaller values of  $d_{sg}$ when $P$ approaches $\lambda_{spp}$. Only a single broad peak lasts when $P>\lambda_{spp}$.

The highest emittance is obtained when $d_{sg}$ is close to $P$ but always smaller than $\lambda_{spp}$; for our choose of parameters $\eta=62$ for $d_{sg}=0.93 P$ and $P=0.95 \: \lambda_{spp}$, see the black line of Fig. \ref{fig:regular}(b). It is worth to stress that a slightly larger P implies a significant reduction of $\eta$. It is commonly assumed  that the ideal groove pitch for extraordinary optical transmission should fulfill SPP constructive interference conditions ($P=n \lambda_{spp}$, where n is an integer), thought it does not occur in practice \cite{GenetN07}; see Ref. \cite{FdLPPRB11} for a discussion in the case of a minimal interactive system. For example, Janssen {\it et al.} obtained an optimal transmittance for $d_{sg}=0.54 \: P$ assuming  $P$ equal to $\lambda_{spp}$  at $\lambda=800$ nm \cite{JanssenPRL07}. However, despite of the disagreement with authors of Ref. \cite{JanssenPRL07} regarding the best choice for $P$, our calculations confirm their results that the transmittance is negligible when $d_{sg}=P=\lambda_{spp}$.

The aspect ratio $r=w_g/h_g$ is another important design parameter. Fig. \ref{fig:regular} (c) shows $\eta$ as a function of $r$ for the optimal geometry of Fig. \ref{fig:regular} (b). It has two peaks centered at $r \approx 1.0$ and $r \approx 6.0$. The position of the second peak changes as a function of the wavelength (not shown). Though $\eta$ is slightly larger for $r=1.0$ than for $r=6.0$ in Fig. \ref{fig:regular} (c), we shall see below that the relative contribution of the two peaks changes when several parameters are optimized simultaneously.


We conclude the analysis of Fig. \ref{fig:regular} stressing that  groove pitch and depth are the most relevant design parameters, leading to $\eta=54$ in Fig. \ref{fig:regular}(a). Ideal values of $r$ and $d_{sg}$ produces an additional enhancement of 15 \%. Up to now we have $\eta=62$ for $\lambda=1.35$ $\mu$m, $h_s=360$ nm, $w_s=100$  nm. $h_g=w_g=125$, $P=0.95 \: \lambda_{spp}$, and $d_{sg}=0.93 P$. This geometry is used below as a seed in the CG optimization. Similar calculations have been done for other wavelengths at optical and IR frequencies (not shown).

\subsection{CG optimization as a function of the wavelength}
\label{subsec:CGvslambda}

We search now for the optimal groove profile scanning the parameter space with help of the conjugate gradient maximization algorithm \cite{NRF77}, that was implemented for that purpose within our CMM formalism.  Optimal geometries as a function of $\lambda$ for a uniform groove array with $N_g=10$ grooves are reported in Table \ref{tab:geom}. 

\begin{table}
\centering
 \begin{tabular}{r|ccccccc|c} \hline \hline 
Parameter & \multicolumn{7}{c}{Gold} & PEC \\ \hline
 $\lambda$ ($\mu$m) & 0.65 & 0.85 & 1.35  & 4.0 & 6.0  & 8.0  & 10.6  & 10.6 \\ \hline
$\lambda_{spp}$ ($\mu$m) & 0.62 & 0.83 & 1.33 & 3.997 & 5.997& 7.998 & 10.599 & -	\\ 
$h_s$ ($\mu$m) & 0.14  & 0.22  & 0.36 & 1.28 & 2.17 & 2.94 & 3.6  & 3.95 \\
  $w_s$ ($\mu$m) & 0.05 & 0.07 & 0.1 & 0.296 & 0.44 & 0.59  & 0.78 & 0.79\\ \hline
  $h_g$ ($\mu$m) & 0.06 & 0.06 & 0.09 & 0.37  & 0.56  & 0.76  & 1.02  & 1.27  \\
$w_g$ ($\mu$m) & 0.38  & 0.4 & 0.45  & 0.94 & 1.47  & 2.06  & 2.68  & 1.65  \\
   $P$ ($\mu$m) & 0.60  & 0.79  & 1.27  & 3.81  & 5.7  & 7.61  & 10.1 & 10.1 \\
$d_{sg}$ ($\mu$m) & 0.53 & 0.72 & 1.17 & 3.49 & 5.27 & 7.028  & 9.3  & 9.35 \\ \hline
$h_s/\lambda$ & 0.22 & 0.26 & 0.27 & 0.32 & 0.36 & 0.37	& 0.37	& 0.37
 \\
$h_g/h_s$ & 0.41 & 0.28 & 0.25 & 0.29 & 0.26 &	0.26 & 0.26 & 0.32 \\
$r=w_g/h_g$ & 6.6 & 6.6 & 4.9 & 2.5 & 2.6 & 	2.7 &	2.6 & 1.3
\\ 
$P/\lambda_{spp}$ & 0.96 & 0.96 & 0.95	& 0.95	& 0.95	& 0.95	& 0.95	& 0.95
\\
 $d_{sg}/P$ & 0.88 & 0.91 & 0.92 &	0.92 & 0.92 & 0.92 & 0.92 & 0.93	\\ \hline
absorption (\%) & 59 & 51 & 41 & 27 & 25 & 23 & 22 & 0\\ \hline
$\eta$         & 35 & 47 & 68 & 77 &  75 & 75 & 74 & 102 \\ 
$\eta$ (SGA scaled by $\lambda$) & - & - & 10 &  55 & 68 & 72 & 74 & - \\ \hline \hline
 \end{tabular}
\caption{Optimal geometries obtained by a CG optimization of a uniform SGA with $N_g=10$ grooves for different values of the wavelength. Last column reports calculations for a PEC at $\lambda=10.6$ $\mu$m. Last row gives the values of $\eta$ for a SGA  scaled by $\lambda$ with respect to the SGA made of gold and optimized at $\lambda=10.6$ $\mu$m.}
\label{tab:geom}
\end{table}

The metal thickness ($h_s$) defining the position of the Fabry-Perot mode for a given wavelength and the slit width ($w_s$) that defines its intensity are keep constant along the simulation process.    For the sake of comparison on an equal footing, we scale $w_s$ at a given $\lambda$ with respect to the value $w_s=100$ nm chosen above for $\lambda=1.35$ $\mu$m. Input values of $h_s$ and $w_s$ are included in Table \ref{tab:geom}.

We observe in Table \ref{tab:geom} that the normalized transmittance for the optimal system increases from $\eta=35$ at $\lambda=0.65$ $\mu$m to $\eta=74$ at $\lambda=10.6$ $\mu$m, due to the reduction of the metal absorption for increasing $\lambda$. The absorption is also given in Table \ref{tab:geom}. It decreases from 59 \% at  $\lambda=0.65$ $\mu$m to 22 \% $\lambda=10.6$ $\mu$m. This quantity is defined as 1-R-T, where R and T are the total reflectance and total transmittance, respectively. 

The most remarkable feature is that, despite the reduction of the absorption, $\eta$ does not increase with $\lambda$ in the mid-IR but states almost constant around the value $\eta=75$. That indicates that radiative losses dominate over dissipative losses. However, Table \ref{tab:geom} shows that dissipative losses are not negligible even in the IR regime. In order to quantify its effect we optimized a SGA in a PEC at the longest wavelength considered in this paper, $\lambda=10.6$ $\mu$m. Results are given in the last column of Table \ref{tab:geom}. We find that, even at the long wavelength part of the IR, $\eta$ for a PEC is 42 \% larger than for gold. A relevant feature of a PEC is that the dimensions of the system can be scaled with $\lambda$, having the resulting structure the same transmittance than the original one. 
We have verified if that simple rule could be applied  to the IR. Our calculations show that it is a nice approximation for the long wavelength part of the IR. In the last row of Table \ref{tab:geom} we report the values of $\eta$ for the SGAs scaled by $\lambda$ with respect to the SGA made of gold and optimized at $\lambda=10.6$ $\mu$m, i.e. we multiply the geometrical parameters of the before-last column of Table \ref{tab:geom} by the scaling factor $\delta_\lambda$, which is equal to $\delta_\lambda=8.0/10.6=0.75$ for $\lambda=8.0$ $\mu$m, $\delta_\lambda=0.57$ for $\lambda=6.0$ $\mu$m, and so forth. A good agreement between the optimal SGA and the scaled one is obtained for $\lambda>6.0$ $\mu$m.

We find however that the ideal $P \approx 0.95 \lambda_{spp}$ and $d_{sg} \approx 0.92 P$ are independent of the wavelength for $\lambda>0.65$ $\mu$m. These values are very close to those obtained for a PEC. This behavior is related to the properties of groove cavity modes \cite{FLTAPA07}. The dispersion relation of surface modes is close to the folded dispersion relation of the flat surface, which is also close to the folded light line (notice that $\lambda_{spp} \approx \lambda$ in Table \ref{tab:geom}). At the center of the Brillouin zone, gaps are open due to the finite size of the grooves. Gaps strongly depend on groove geometry and wavelength, leading to an ideal aspect radio that decreases from 6.6 to 2.6 for increasing $\lambda$, as well as an optimal groove depth that oscillates around a quarter of the metal thickness. Geometrical changes are larger for a PEC due to the different scattering cross-sections of real and perfect electric conductors.

\begin{table}
\centering
 \begin{tabular}{c|ccccccc|c} \hline \hline 
Parameter & \multicolumn{7}{c}{Gold} & PEC \\ \hline
$\lambda$ ($\mu$m) & 0.65 & 0.85 & 1.35 & 4.0 & 6.0  & 8.0 & 10.6 & 10.6 \\ \hline
 $\Delta \lambda$ ($\mu$m) & 0.04 & 0.05 & 0.06 & 0.17 & 0.30 & 0.42 & 0.56 & 0.48 \\ 
$\Delta h_s$ ($\mu$m) & 0.13 & 0.26 & 0.23 & 0.63 & 0.90 & 1.31 & 1.65 & 1.46\\
  $\Delta h_g$ ($\mu$m) & 0.08 & 0.05 & 0.06 & 0.18 & 0.31 & 0.43 & 0.57 & 0.36 \\
$\Delta w_g$ ($\mu$m) & 0.20 & 0.27 & 0.65 & 1.96 & 3.02 & 4.02 & 5.05 & 1.29 \\
   $\Delta P$ ($\mu$m) & 0.04 & 0.05 & 0.075 & 0.23 & 0.36 & 0.50 & 0.67 & 0.56\\
$\Delta d_{sg}$ ($\mu$m) & 0.16 & 0.18 & 0.20 & 0.61 & 1.21 & 1.65 & 2.25 & 1.83\\ \hline \hline
 \end{tabular}
\caption{FWHM of $\eta$ as a function of each parameter reported in \ref{tab:geom}. In particular, the FWHM for $\eta$ versus $\lambda$ is the bandwidth. Last column reports calculations for a PEC at $\lambda=10.6$ $\mu$m.}
\label{tab:fwhm}
\end{table}

 The tolerance of the optimal system to small changes of its geometry is estimated by the full width at half maximum (FWHM) of $\eta$ as a function of the corresponding geometrical parameter. The FWHM for the optimal geometry given in Table \ref{tab:geom} is reported in Table \ref{tab:fwhm}. In particular, the FWHM for $\eta$ versus $\lambda$ is the bandwidth. We find that the tolerance on the period is smaller than for $d_{sg}$. The FWHM is also larger for  $w_g$ than for $h_g$. The reason for this behavior is that period and groove depth give rise to the largest enhancement of $\eta$ when groove cavity modes are excited. So, peaks are narrower when the system is more resonant, as one should expect.  Interestingly, $\eta$ versus $w_g$ has a broad peak (not shown) instead of the two peaks depicted in Fig. \ref{fig:regular}(c).

\subsection{CG optimization as a function of the system size}
\label{subsec:CGvsNg}
\begin{figure}[htbp]
 \centering
\includegraphics[width=6cm]{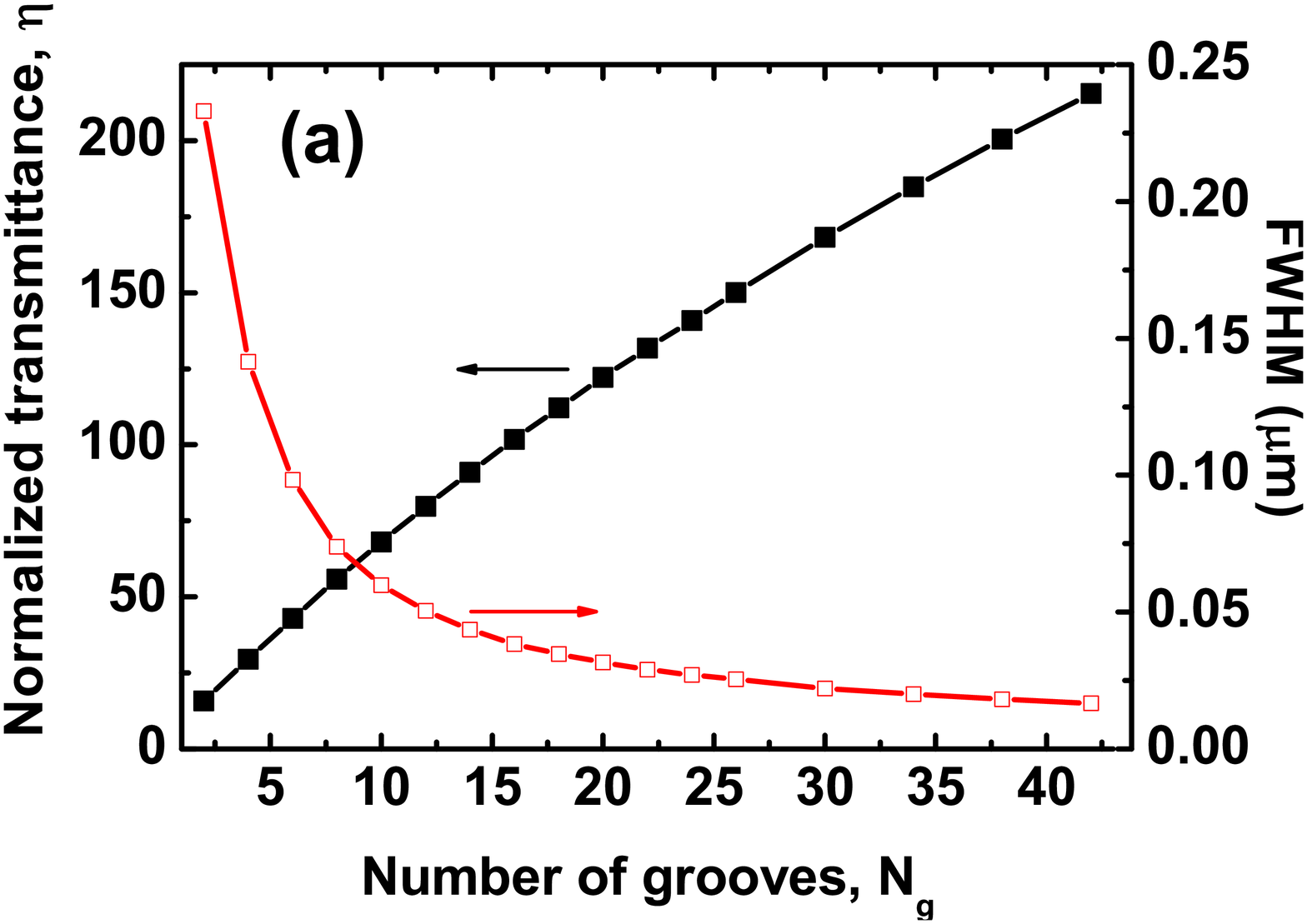}
\includegraphics[width=6cm]{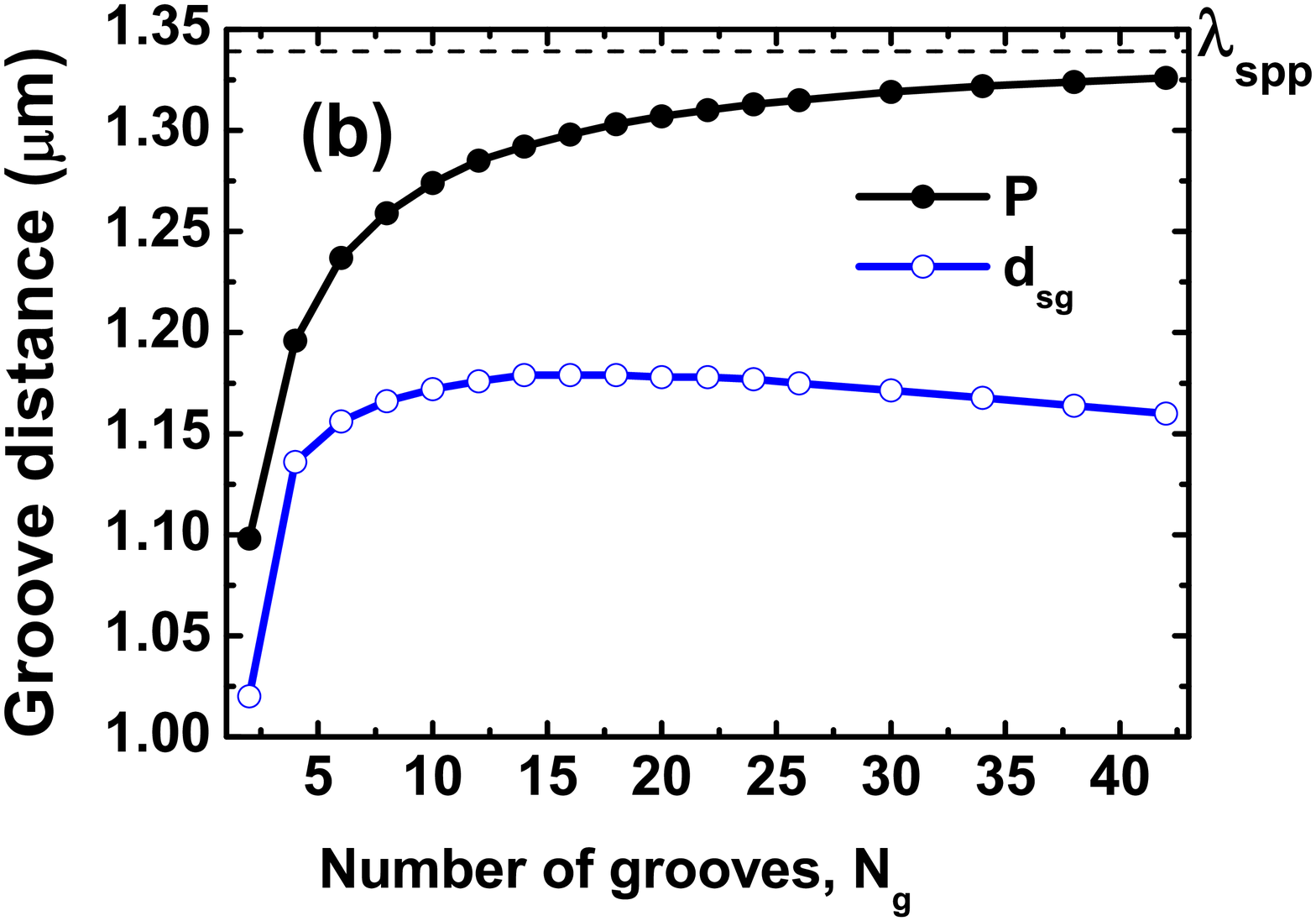}
\includegraphics[width=6cm]{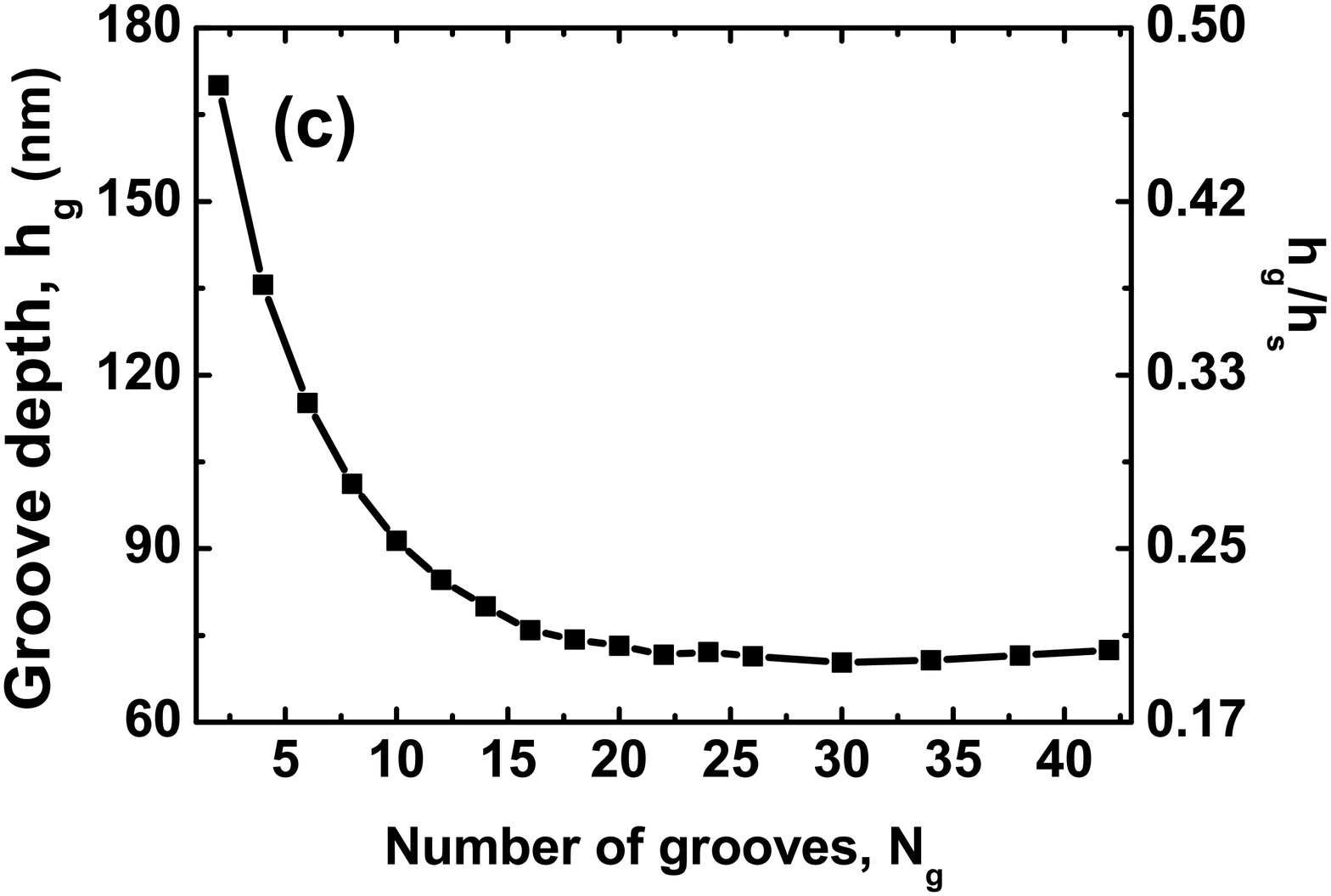}
\includegraphics[width=6cm]{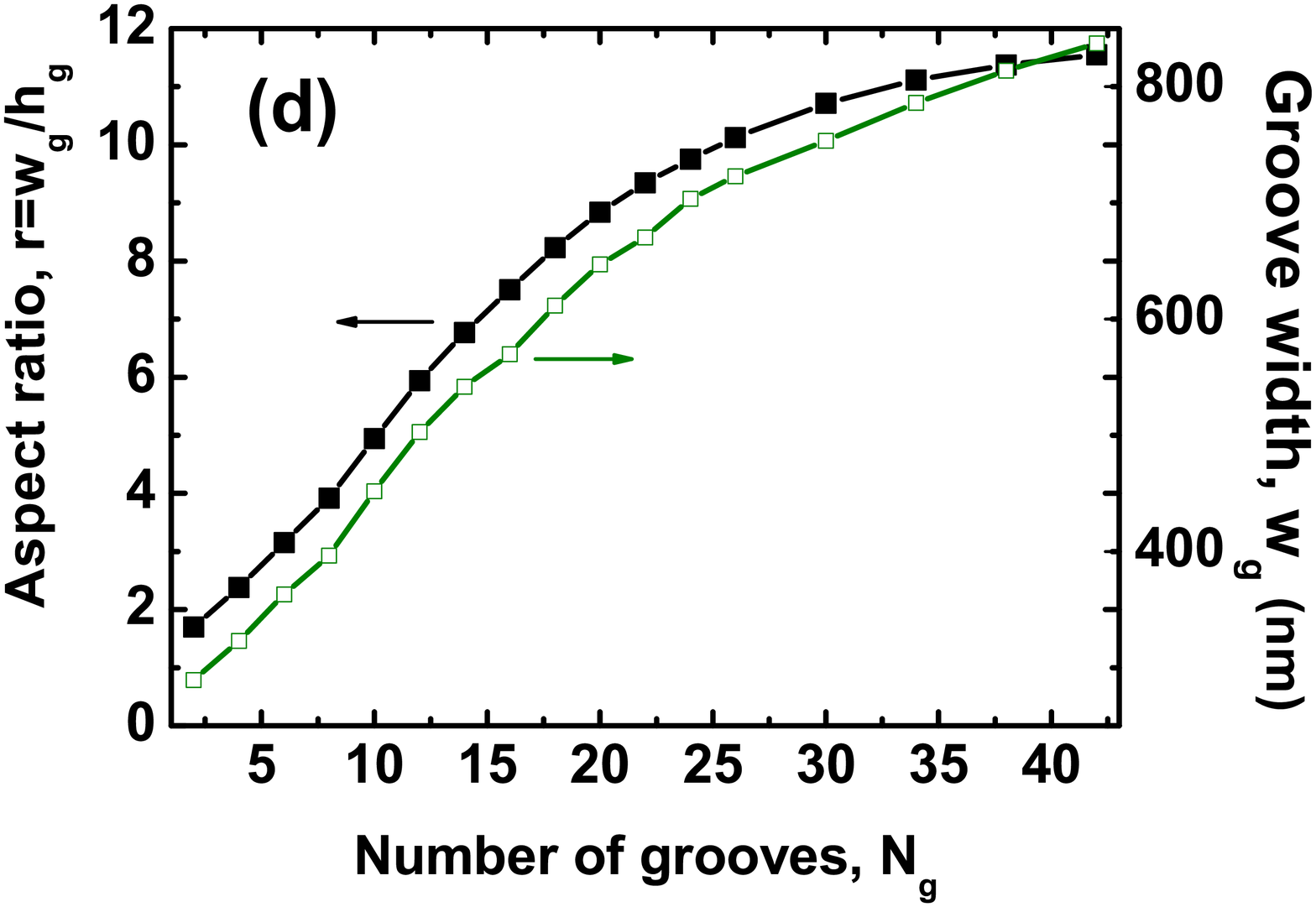}
\caption{(Color). CG optimization of a SGA for $\lambda=1.35$ $\mu$m and $N_g$ increasing from 2 to 42. (a) Normalized-to-area transmittance, $\eta$, and FWHM. (b) Period, $P$, and slit-nearest groove distance, $d_{sg}$. (c) Groove depth, $h_g$. (d) Aspect radio, $r=w_g/h_g$, and groove width, $w_g$.}
 \label{fig:vsNg}
\end{figure}

In this section we optimized a uniform SGA as a function of the number of grooves, $N_g$. A CG maximization is performed for a given value of $N_g$ and $\lambda$ using as a seed the geometries reported in Table \ref{tab:geom}. Fig. \ref{fig:vsNg} shows our results for $N_g$ increasing from 2 to 42 and $\lambda=1.35$ $\mu$m.

We observe in Fig. \ref{fig:vsNg}(a) that the normalized-to-area transmittance grows monotonically up to $\eta=216$ for $N_g=42$.
As pointed out in Ref. \cite{JanssenPRL07}, in order to achieve such enhancement the distance between grooves raises as a function of $N_g$ and the groove depth decreases, see Figs. \ref{fig:vsNg} (b) and (c). In particular, we find that $P$ approaches $\lambda_{spp}$ for increasing $N_g$. In contrast, $d_{sg}$ reaches its maximum value for only 14 grooves. It means that the central slit no longer interacts with the outer grooves when $N_g$ is larger than this value. On the other hand, $h_g$ decreases by 100 nm as a function of $N_g$ until it reaches the value $h_g=72$ nm for $N_g=22$. Then, $h_g$ remains practically constant. We add to the conclusions of Ref. \cite{JanssenPRL07} that, in order to perform the optimal LHP when more grooves are added to the SGA, grooves should be not only shallower but also broader, see Fig. \ref{fig:vsNg} (d), where the aspect ratio raises until $r=11.6$. Notice that  only $w_g$ still grows for $N_g>20$ whereas $P$, $d_{sg}$, and $h_g$ are practically constant. We also see in Fig \ref{fig:vsNg}(a) that the FWHM decreases with $N_g$. Thus, the price for the enhancement of $\eta$ is having a more resonant SGA. 	

Features found in Fig. \ref{fig:vsNg} can be explained as follows. Given that the light harvesting process is mainly dominated by radiation losses, the SGA increases its total size to collect additional light, thus increasing $P$ and $d_{sg}$. For the same reason, broader grooves enhance the scattering cross-section between modes inside the grooves and the EM radiation \cite{FLTPRB05}.  On the other hand, as the radiation pattern is driven by surface modes running along the metal/air interface, the grooves tend to be shallower to reduce the shadows that apertures closer to the central slit produce on apertures on the edge of the SGA. Shallower grooves also implies a smaller scattering cross-section, but this effect is compensated with broader grooves and a larger system size.

\section{Nonuniform slit-groove arrays}
\label{sec:nonregular}

We can also vary the geometry and pitch of all the grooves, in such a way that the optimized groove array becomes aperiodic and the groove geometry change with the groove position.  In such a case, physical intuition can not longer be used and the optimization process is based solely on the CG algorithm. The optimization routine is divided into the following steps:

\noindent {\em i) Optimization of a uniform SGA:} already discussed in the previous sections.

\noindent {\em ii) One-parameter optimization:} A given parameter is let to vary	 for each groove, so the groove array is no more uniform. Independent simulations are run for the set of groove depths, $\left\lbrace h_g \right\rbrace$, groove widths, $\left\lbrace w_g \right\rbrace$, and groove positions, $\left\lbrace x_g \right\rbrace$.  

\noindent {\em iii) Two-parameter optimization:} A pair of parameters is let to vary	 for each groove. Independent simulations are run for the sets: $\left\lbrace h_g, x_g \right\rbrace$, $\left\lbrace h_g, w_g \right\rbrace$, and  $\left\lbrace w_g,x_g \right\rbrace$. 

\noindent {\em iv) Full groove geometry, $\left\lbrace h_g, w_g, x_g \right\rbrace$:} All geometrical parameters in each groove are let to vary. 

\begin{figure}[htbp]
 \centering
\includegraphics[width=6cm]{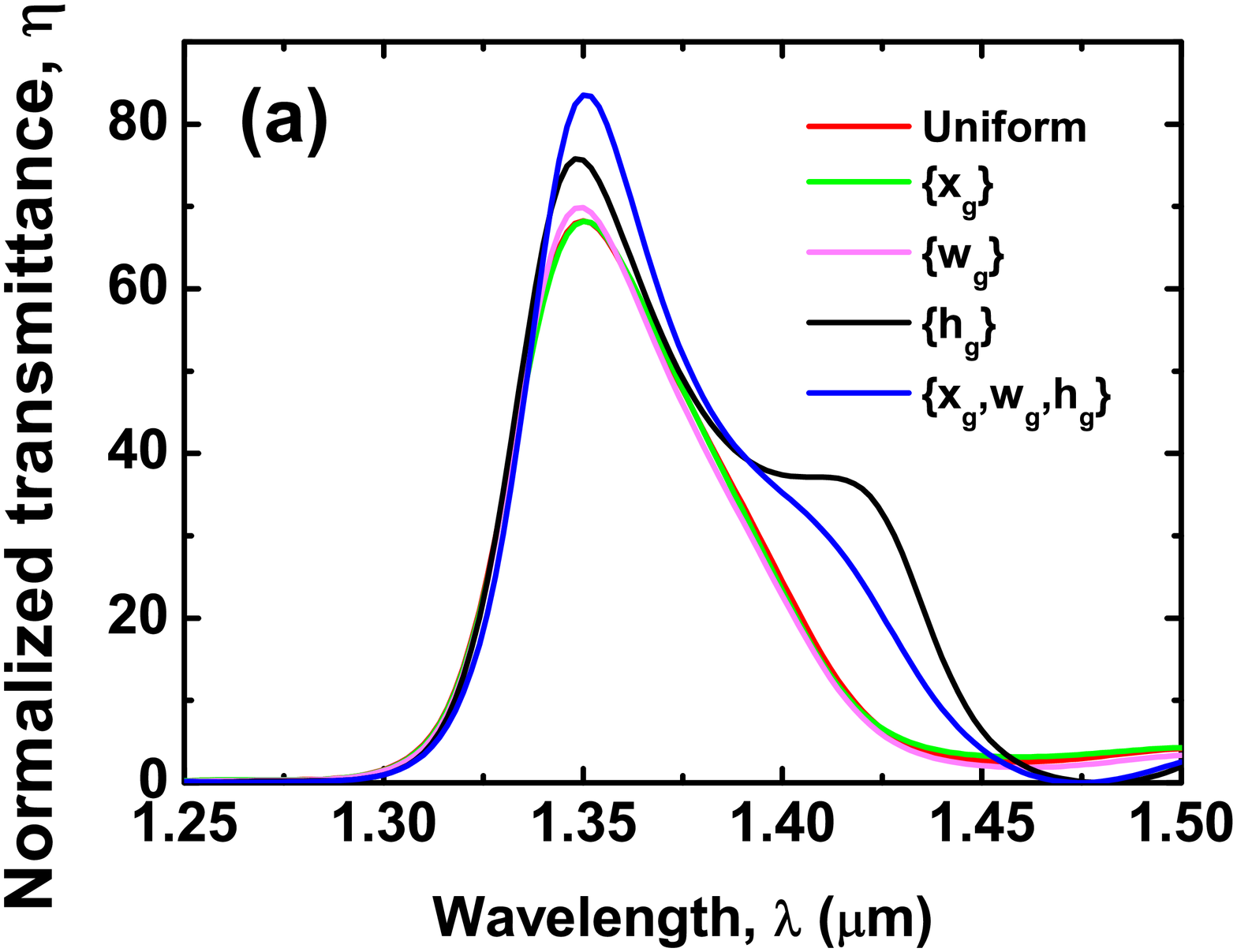}
\includegraphics[width=6cm]{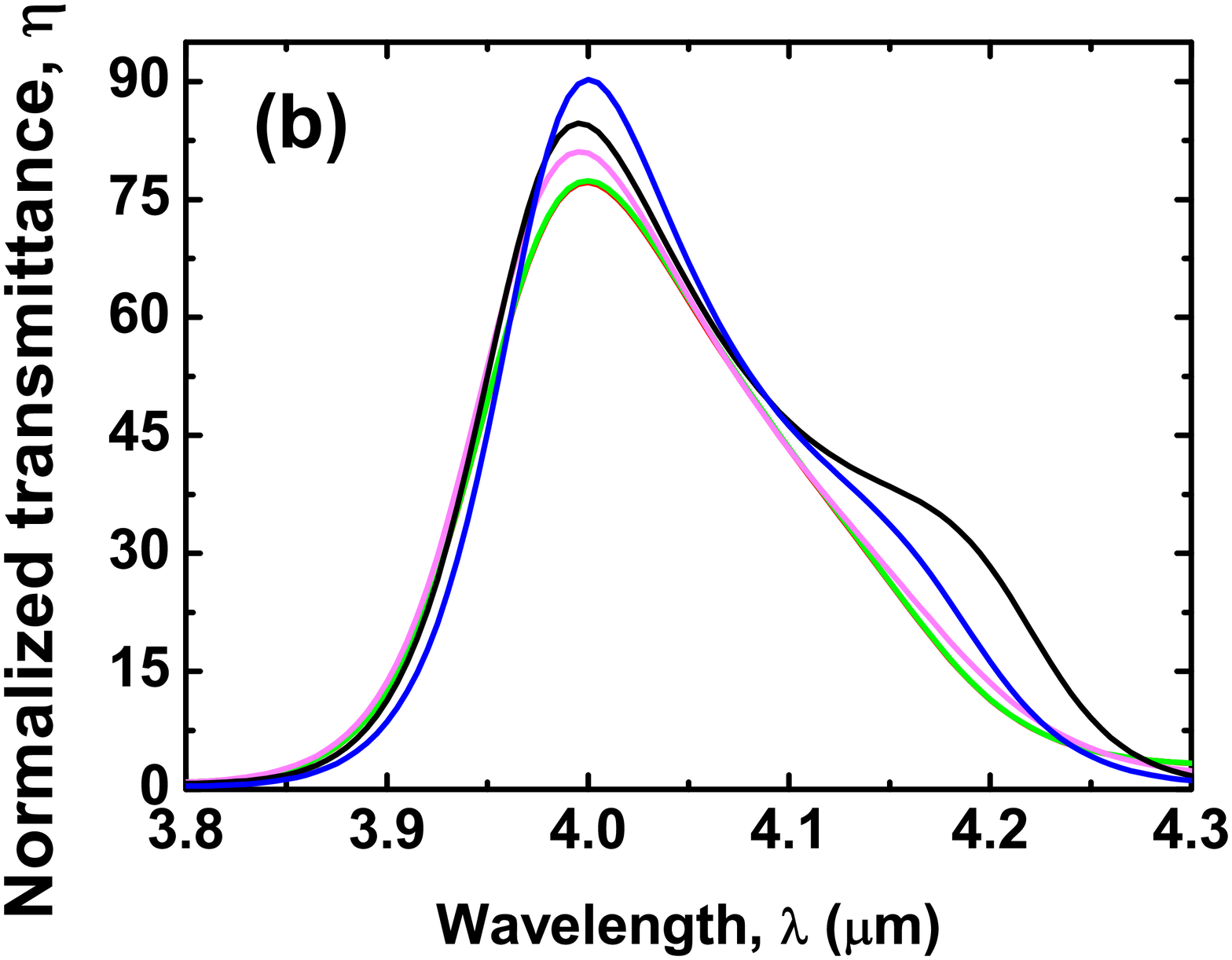}
\includegraphics[width=6cm]{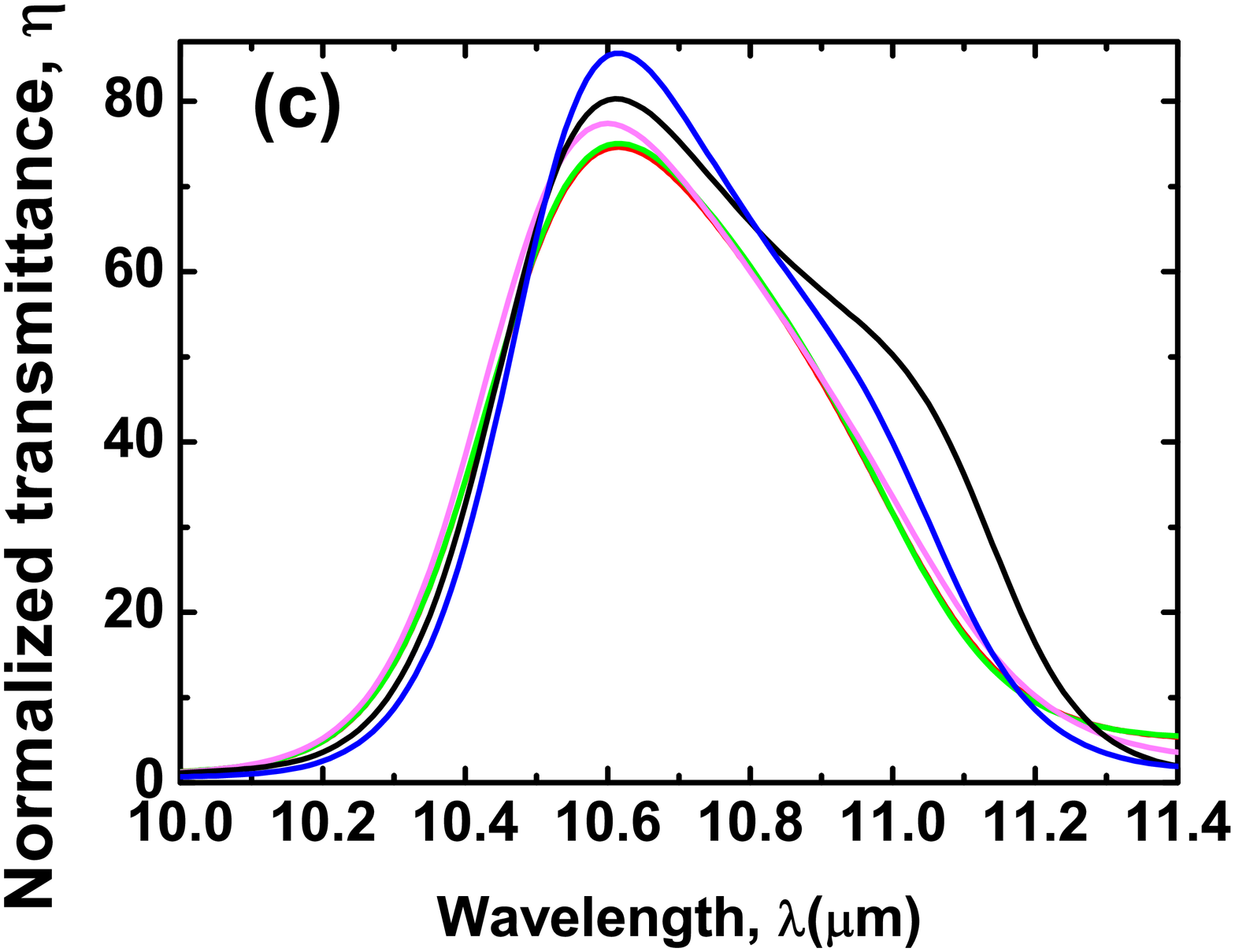}
\caption{(Color). Spectra of the SGA optimized with the CG algorithm for the target wavelengths (a) $\lambda=1.35$ $\mu$m, (b) $\lambda=4.0$ $\mu$m, and (c) $\lambda=10.6$ $\mu$m. Full $\left\lbrace h_g, w_g, x_g \right\rbrace$ and one-parameter optimizations ($\left\lbrace h_g \right\rbrace$, $\left\lbrace w_g \right\rbrace$, and $\left\lbrace x_g \right\rbrace$) are compared with the optimization for a uniform SGA.}
 \label{fig:nonregular}
\end{figure}

\begin{figure}[htbp]
 \centering
\includegraphics[width=6cm]{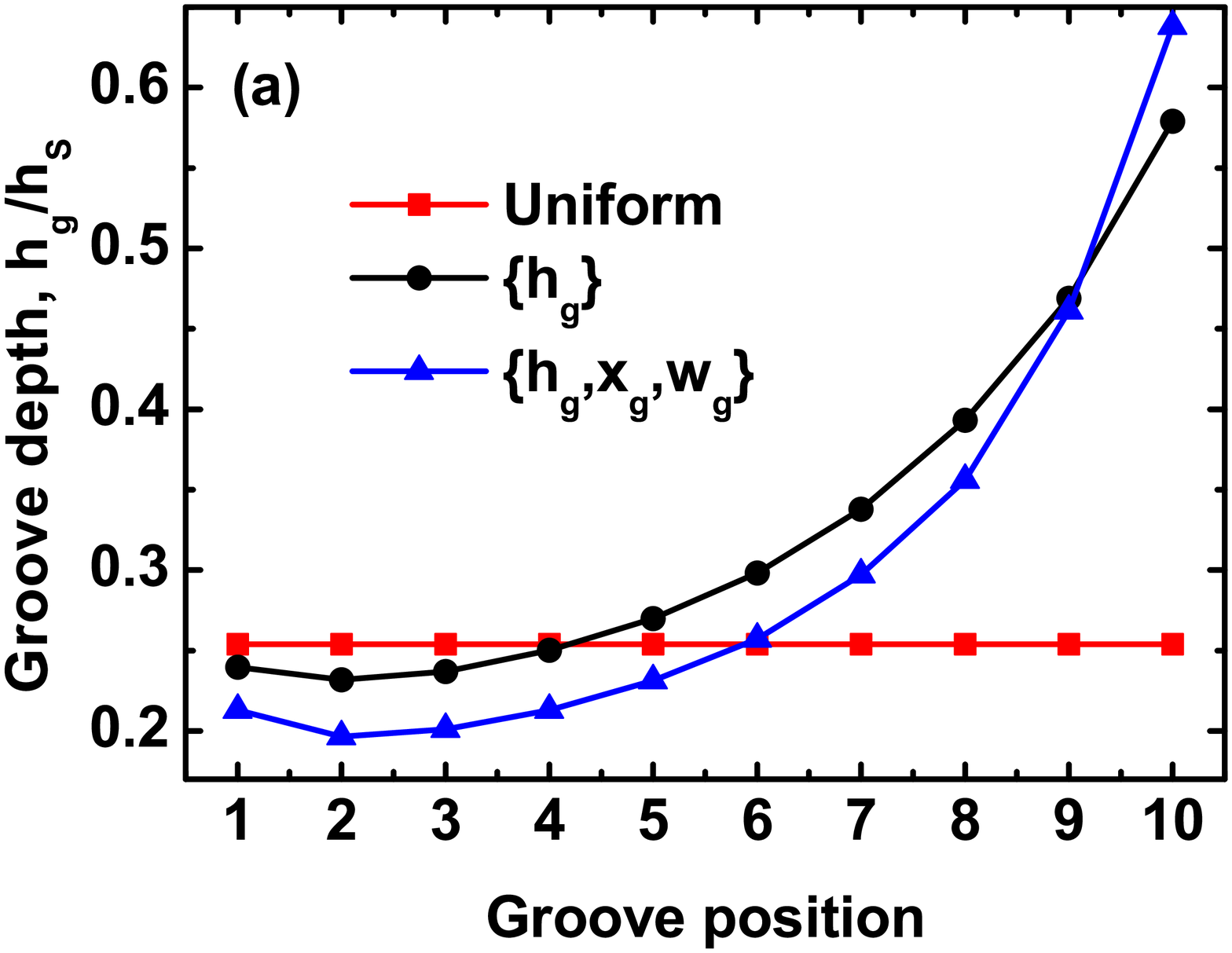}
\includegraphics[width=6cm]{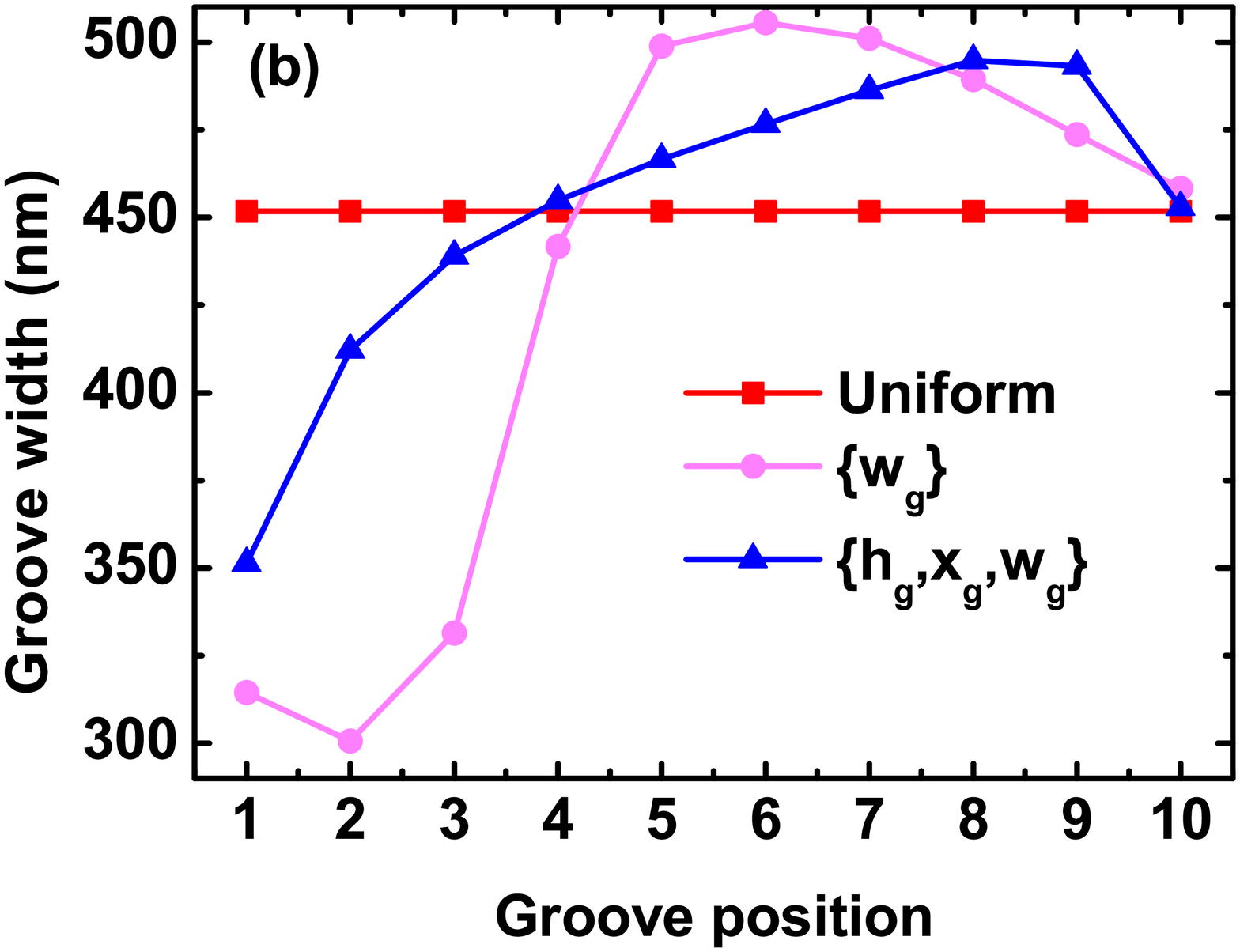}
\includegraphics[width=6cm]{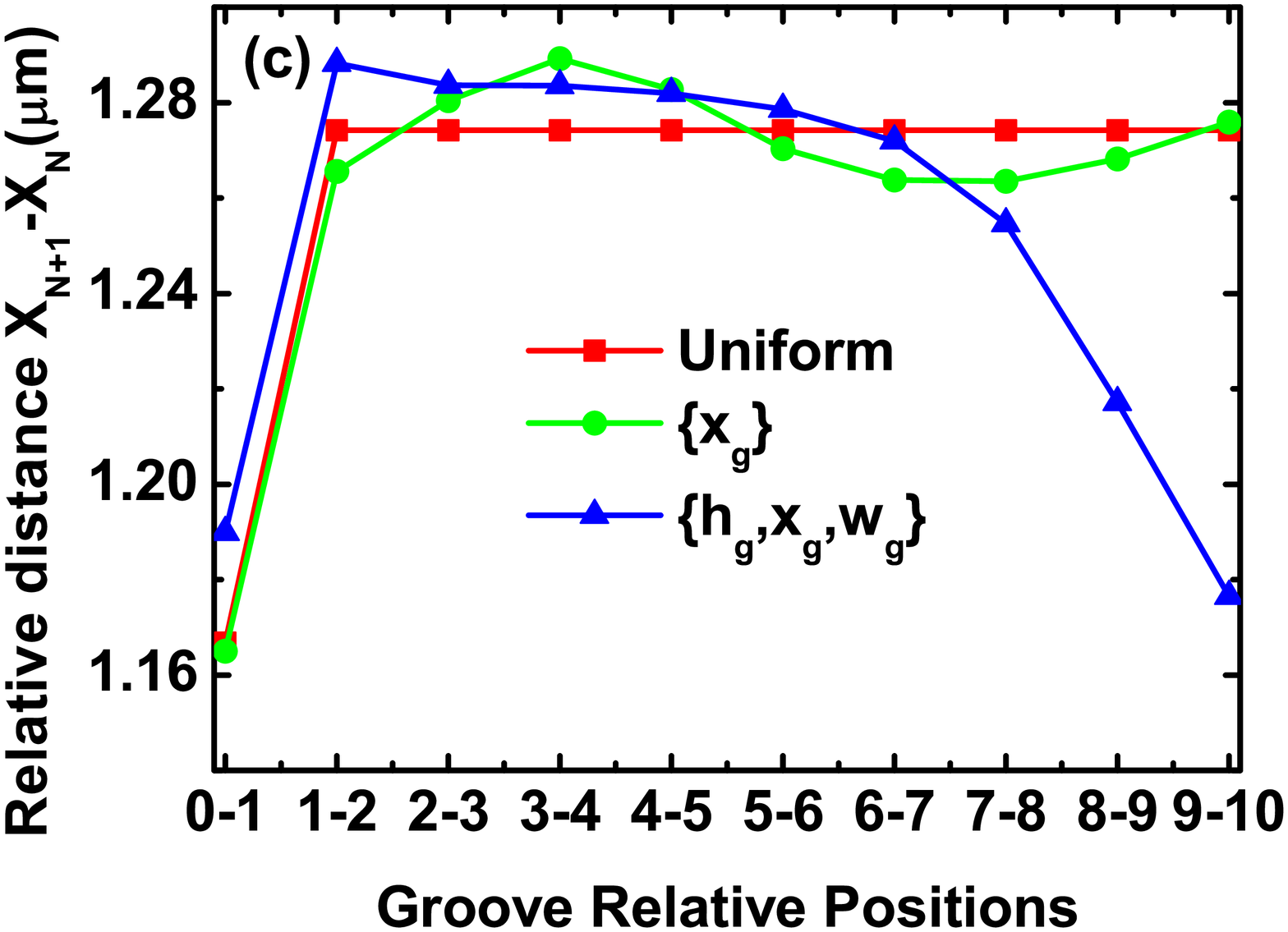}
\includegraphics[width=6cm]{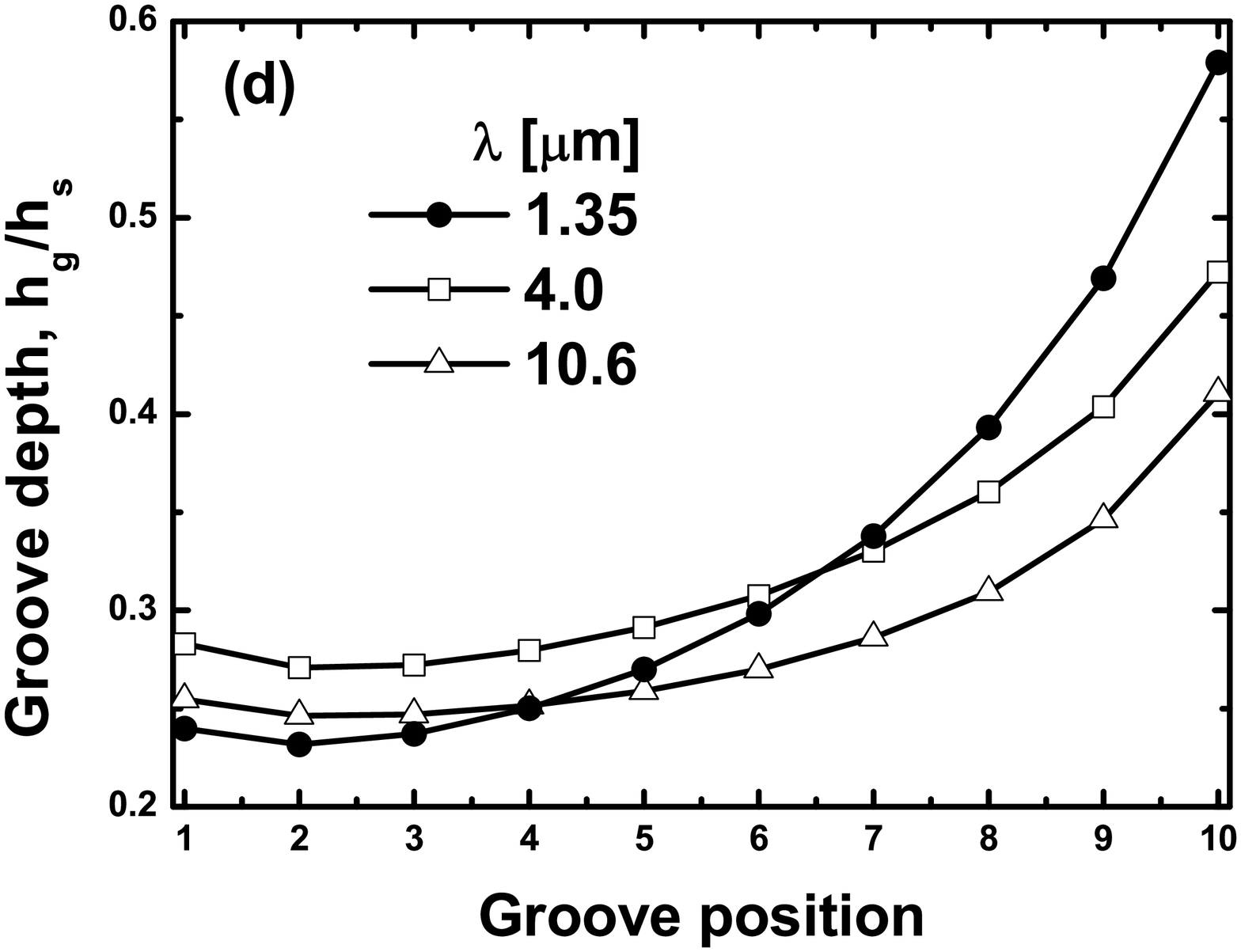}
\caption{(Color). Profiles obtained  for $\lambda=1.35$ $\mu$m in the CG optimizations of the uniform and nonuniform SGAs. (a) Profiles of groove depths. (b) Profiles of groove widths. (c) Profile of groove positions. (d) Groove profiles as a function of $\lambda$.}
 \label{fig:profiles}
\end{figure}

Optimal geometries obtained in one step are used as a seed for next step. Several trials with different seeds  have been done for checking the consistency of the numerical results. We find that the variation of each parameter is within the tolerances reported in Table \ref{tab:fwhm}. Spectra for SGAs optimized at several $\lambda$s are represented in Fig. \ref{fig:nonregular}. Typical profiles are depicted in Fig. \ref{fig:profiles}. 

We find that  the enhancement provided by a nonuniform SGA (optimizing the full groove geometry) with respect to a uniform one is  39\% at $\lambda=1.35$ $\mu$m, 17 \% at $\lambda=4.0$ $\mu$m, and 15\% at  $\lambda=10.6$ $\mu$m, see Fig. \ref{fig:nonregular}. The enhancement decreases with the wavelength.

The variation of the groove depth provides optimal geometries among one-parameter optimizations, with an enhancement of 11 \% for $\lambda=1.35$ $\mu$m, 10 \% for $\lambda=4.0$ $\mu$m, and 8\% for  $\lambda=10.6$ $\mu$m. Fig. \ref{fig:profiles}(a) shows the groove profiles obtained at $\lambda=1.35$ $\mu$m after the optimization of a uniform SGA (red), the one-parameter optimization $\left\lbrace h_g \right\rbrace$ (black), and the full-parameter optimization $\left\lbrace h_g, w_g, x_g \right\rbrace$ (blue). The optimal groove-depth profile for $\left\lbrace h_g \right\rbrace$ is a  chirped array with groove that deepens from $0.24 h_s$ to $0.58 h_s$. A similar variation between $0.21 h_s$ to $0.64 h_s$ is observed for the full parameter optimization.  We also find that grooves occupy a smaller fraction of the metal thickness for longer wavelengths, see Fig. \ref{fig:profiles}(d). In all these cases, distant grooves need to be deeper than those close to the central aperture in order to increases their scattering cross section and squeeze additional light into the aperture.  



The optimal width profile is also a  chirped array with incremental groove width, see Fig. \ref{fig:profiles}(b). As for the depth profile, distant grooves are broader for they should increase their scattering cross section in order to further enhance the transmittance through the central aperture. We also observe a difference between $\left\lbrace w_g \right\rbrace$ and $\left\lbrace h_g, w_g, x_g \right\rbrace$ optimizations larger than for the groove profile. The optimal geometry is consistent with previous calculations that found an  efficiency of 50\% in plasmon generation using a grating with a gradient in the groove width \cite{LuJOSAB07}. 

We do not find large deviation from the uniform SGA when the groove-position profile is optimized with the CG algorithm, see Fig. \ref{fig:profiles}(c). The relative distance between grooves is practically the same than for a uniform SGA, except for the farthest groove of the array in the full-parameter optimization. The optimization of the groove-position profile  does not produce an appreciable enhancement of the normalized transmittance with respect to the uniform SGA, see Fig. \ref{fig:nonregular}.

\section{Conclusions}

In search for high efficiencies in the light harvesting process of a one-dimensional slit-groove array, we have considered the interplay between Fabry-Perot modes of a single slit and groove cavity modes of the groove array. We have developed the following simple design rules for a uniform SGA, that are valid for optical and IR frequencies.
\begin{itemize}
\item Fabry-Perot and groove cavity modes should be at the same spectral position. For subwavelength apertures at a given $\lambda$, the position and intensity of the Fabry-Perot is controlled by metal thickness and aperture size, respectively. The spectral position of the groove cavity modes is mainly determined by groove depth and pitch.
\item The normalized-to-area transmittance grows monotonically with the number of grooves, given that grooves move apart and become broader and shallower as $N_g$ increases. In particular:
\begin{itemize}
\item The ideal periodicity (P) is slightly smaller than the SPP wavelength ($\lambda_{spp}$). $P$ tends to $\lambda_{spp}$ (from below) when more grooves are added to the SGA.
\item Slit-nearest groove distance should be about $0.92 P$ (but only for the optimal P).
\item  The ideal groove depth is smaller than half of the metal thickness. This quantity decreases with the number of grooves ($N_g$).
\item The optimal aspect ratio decreases with $\lambda$ but raises with $N_g$.
\end{itemize}
\end{itemize}

We also account the following trends for the optimal system:

\begin{itemize}
 \item The transmission efficiency of a uniform SGA in the infrared is practically independent of $\lambda$. 
 \item In contrast, the enhancement provided by a nonuniform SGA decreases with $\lambda$. A chirped groove array enhances the transmittance between 15\% and 39\% for decreasing $\lambda$.
 \item The bandwidth decreases with the size of the system.	
\end{itemize}
Summing it up, we hope that our findings could motivate further theoretical and experimental studies of light harvesting structures.

\section*{Acknowledgments}
The authors gratefully acknowledge financial support by European Projects FP6 PLEAS project Ref. 034506 and EC FP7-ICT PLAISIR Project  Ref. 247991, and the Spanish Ministry of Science and Innovation project MAT2011-28581-C02-02.

\end{document}